\documentclass{jfm}
\usepackage{graphicx}
\usepackage{epstopdf, epsfig}
\usepackage{natbib}
\usepackage{amsmath}
\usepackage{mathtools}
\usepackage{subcaption} 
\usepackage{cancel} 
\usepackage{hyperref}
\usepackage{comment}
\usepackage{tikz}
\hypersetup{
    colorlinks=true,
    linkcolor=blue,
    citecolor=blue
    }
\usepackage{xcolor}
\usepackage{float}

\shorttitle{Pressure-gradient boundary layers}
\shortauthor{V. Baxerres, R. Vinuesa and H. Nagib}

\title{Evidence of quasi equilibrium in pressure-gradient turbulent boundary layers}

\author{Victor Baxerres\aff{1},
  Ricardo Vinuesa\aff{2}
 \and
  Hassan Nagib\aff{1}\corresp{\email{nagib@iit.edu}}}

\affiliation{

\aff{1}ILLINOIS TECH (IIT), Chicago, IL, 60616, USA
\aff{2}FLOW, Engineering Mechanics, KTH Royal Institute of Technology, Stockholm, Sweden
}

\begin{document}

\maketitle

\begin{abstract}
Two sets of measurements utilizing hot-wire anemometry and oil film interferometry for flat-plate turbulent boundary layers, exposed to various controlled adverse and favorable pressure gradients, are used to evaluate history effects of the imposed and varying freestream gradients.  The results are from the NDF wind tunnel at ILLINOIS TECH (IIT) and the MTL wind tunnel at KTH, over the range $800 < Re_\tau < 22,000$ (where $Re_{\tau}$ is the friction Reynolds number).  The streamwise pressure-gradient parameter $\beta \equiv (-\ell/\tau_{w}) \cdot (\partial P_{e}/\partial x)$ varied between $-2 < \beta < 7$, where $\ell$ is an outer length scale for boundary layers equivalent to the half height of channel flow and the radius of pipe flow, and is estimated for each boundary-layer profile. Extracting from each profile the three parameters of the overlap region, following the recent work of \cite{mon23} that led to an overlap region of combined logarithmic and linear parts, we find minimum history effects in the overlap region. Thus, the overlap region in this range of pressure-gradient boundary layers appears to be in ``quasi equilibrium".  
\end{abstract}


\section{Introduction}

A topic within the study of wall-bounded turbulence that has received extensive attention for many decades, due to its significant implications on flow representation and modeling, is the overlap region between the inner or wall region and the outer region of the flow.  According to the classical literature, the mean velocity profile in this overlap region follows the well-known logarithmic law (\ref{eq:001}), which contains $\kappa$, the von K\'arm\'an constant. 
\begin{equation}\label{eq:001}
    \overline{U}^+_{x\rm OL}(y^+\gg 1~\&~Y\ll 1) = \frac{1}{\kappa} \ln y^+ + B.
\end{equation}
In this equation, $\overline{U}_x$ is the streamwise mean velocity, the ``${\rm +}$" superscript indicates that we are using the viscous scaling, and the ``${\rm OL}$" subscript indicates validity in overlap region. The independent variable $y^+$ is the inner-scaled wall-normal coordinate, $y^+ = y u_\tau /\nu$, where $u_\tau$ is the friction velocity and $\nu$ is the fluid kinematic viscosity. The variable $Y$ is the wall-normal coordinate made dimensionless by the ``outer scale of the flow" $\ell$. 
Since the work of \cite{variations} (CN08 from now on) the universality of the log law and the von K\'arm\'an ``coefficient" have been occasionally challenged or reaffirmed, {\it e.g.}, see references by~\cite{george,nagib_debate,vinuesa_exp,luchini}. More recently, \cite{mon23} (MN23 from now on) shed additional light on this topic, challenging the accumulated knowledge during the last century. The main point of MN23 is that a more thorough matched asymptotic expansion (MAE) of wall-bounded flows yields an overlap region containing a combination of logarithmic and linear terms of the distance to the wall in (\ref{eq:010}). They arrived at this result by considering in the inner asymptotic expansion a term proportional to the wall-normal coordinate, $\mathcal{O}(Re_{\tau}^{-1})$. Here we will reference this overlap region as ``log+lin", and use $Re_{\tau} = u_\tau \ell / \nu$, which is the friction Reynolds number with $\ell$ a characteristic outer length. For consistency with MN23, this $\ell$ can be thought of as the radius $R$ in pipes, the semi-height in channels, $h$, or a length scale related to the 99\% boundary-layer thickness $\delta_{99}$ in boundary layers. 
\begin{equation}\label{eq:010}
\overline{U}^+_{x\rm OL}(y^+\gg 1~\&~Y\ll 1) = \kappa^{-1}\ln y^+ + S_0  y^+ /Re_\tau + B_0 + B_1/Re_\tau.
\end{equation}
Earlier publications by \cite{yaj70,afzal73,lee15,luchini} have considered a linear term in the overlap region of channel and pipe flows. However, unlike MN23, none of them revealed the non-universality of the K\'arm\'an constant, or the dependence of the overlap coefficients $\kappa$, $S_0$ and $B_0$ on the pressure gradient along the flow.  In most literature on boundary layers, including for turbulent conditions, the commonly used pressure-gradient parameter is $\beta^{*} \equiv (-\delta^{*}/\tau_{w}) \cdot (\partial P_{e}/\partial x)$, where $\delta^{*}$ is the displacement thickness, and $P_{e}$ is the freestream static pressure.  In order to compare results from developing boundary layers to fully developed flow in channels and pipes, at corresponding pressure gradients, we will use the parameter $\beta \equiv (-\ell/\tau_{w}) \cdot (\partial P_{e}/\partial x)$.  The outer length scale of all boundary-layer profiles used here, $\ell$, is established in the next section.  Effects of pressure gradient on boundary layers have often been considered to have ``history effects" on the flow. In particular, several studies have documented flow-history effects on the Reynolds stresses and the wake region of the mean velocity profile~\citep{harun,bobke}, while other studies have also reported an effect of flow history on the overlap region and its coefficients $\kappa$ and $B$~\citep{Devenport}.

Our objective here is to examine whether such history effects are present in the overlap region, or a behavior more like quasi equilibrium exists when the log+lin overlap is used to analyse well-documented boundary-layer data under conditions of adverse pressure gradient (APG) or favorable pressure gradient (FPG), including strong favorable pressure gradient (SFPG), and for reference, zero pressure gradient (ZPG).

\section{Boundary-layer databases}
Two sets of boundary-layer data were selected and used because the pressure gradients imposed were well controlled and documented, and independent wall-shear stress was measured directly by oil film interferometry. One was carried out in the NDF wind tunnel at ILLINOIS TECH (I.I.T., Chicago)  (\cite{iutam100}) and the other was obtained in the MTL wind tunnel of KTH (Stockholm) (\cite{sanm20}). Both sets of data were measured by hot-wire anemometry and the free stream velocity was measured independently using Pitot probe measurements at reference points above the boundary layers developing on a suspended flat plate. In the case of the MTL data we did not use the "strongly increasing" pressure gradient data, and only the 30 m/s cases to avoid any low $Re_\tau$ effects and insure fully turbulent conditions even near the start of the boundary layer development. More cases were studied by \cite{sanm20}.

\begin{figure}
    \centering
    \begin{minipage}[t]{0.495\textwidth}
        \centering
        \includegraphics[width=\linewidth]{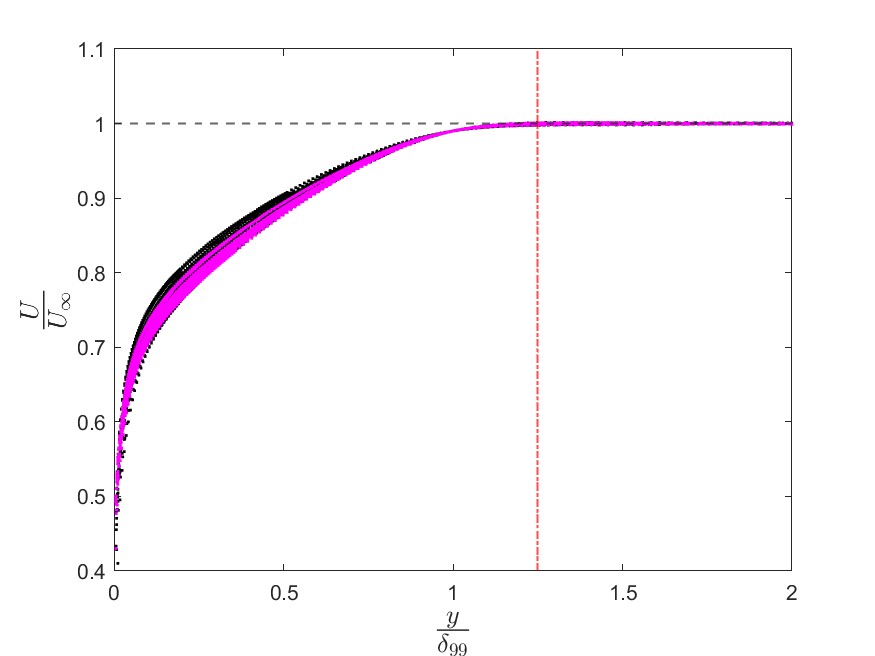}
        \begin{picture}(0,0)
            \put(-20,35){\includegraphics[width=0.5\textwidth]{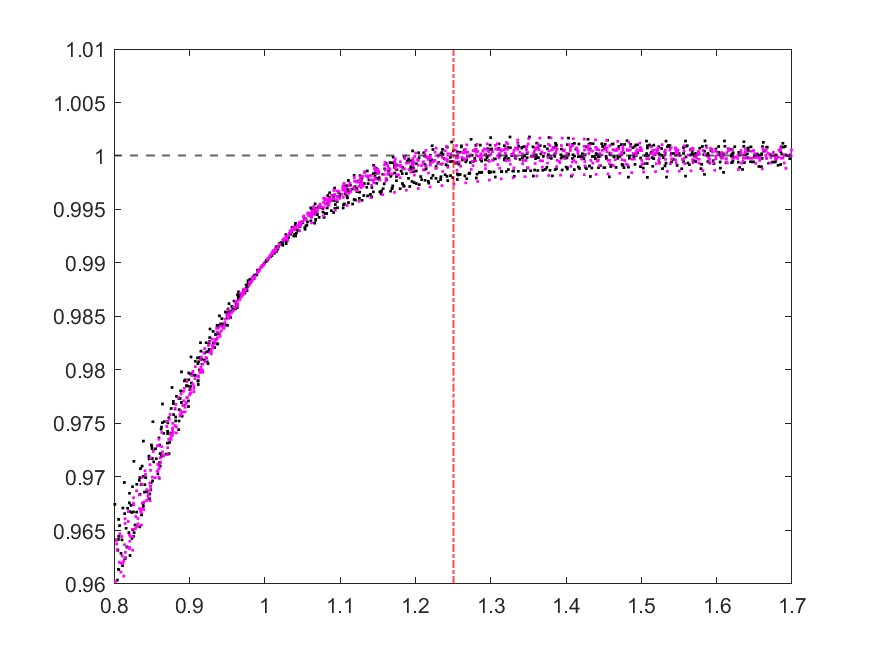}} 
        \end{picture}
        \label{fig:FPG}
    \end{minipage}
    \hfill
    \begin{minipage}[t]{0.495\textwidth}
        \centering
        \includegraphics[width=\linewidth]{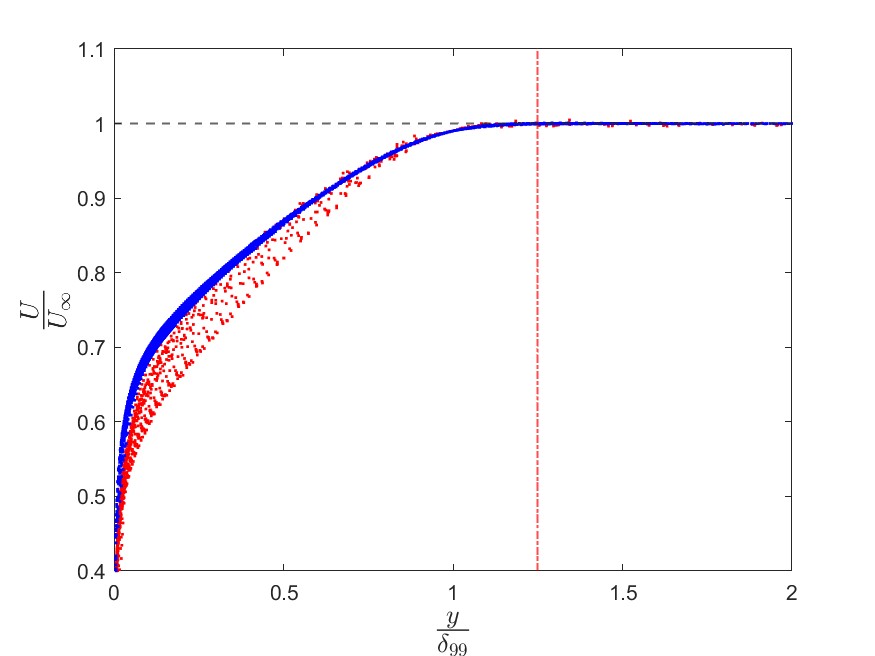}
        \begin{picture}(0,0)
            \put(-20,35){\includegraphics[width=0.5\textwidth]{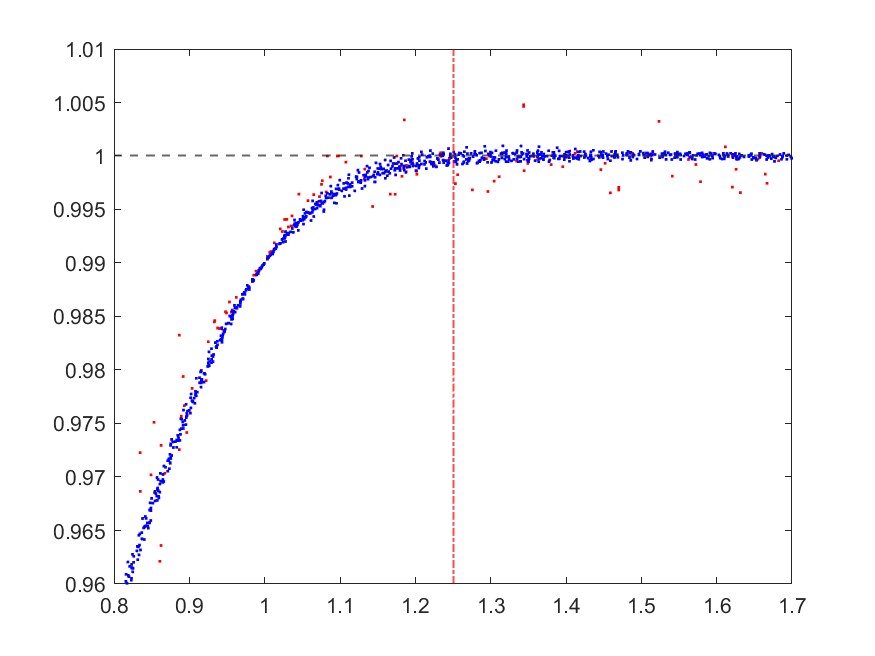}} 
        \end{picture}
        \label{fig:APG}
    \end{minipage}
    \caption{Mean velocity profiles from boundary layers of data sets used (\cite{iutam100,sanm20}, normalized by free stream velocity  $U_{\infty}$ and $\delta_{99}$; (Left) FPG in magenta and SFPG in black from NDF; (Right) APG in red from MTL and APG in blue from NDF in blue; dashed red lines at $y/\delta_{99} = 1.25$, corresponding to $\Delta_{1.25}$.}
 \label{fig:fig1}
 \end{figure}

The velocity profiles of all the data used by us are shown in figure~\ref{fig:fig1}. Since the work by \cite{FDR09}, a composite mean velocity profile has been used to determine an outer scale of ZPG boundary layers instead of various other measures of the boundary-layer thickness, including integral thicknesses, {\it e.g.}, \cite{sam18} and MN23.  This outer scale $\delta$ was systematically compared with the large data base used by \cite{FDR09}, and we concluded that for ZPG, $\delta \approx 1.25 \delta_{99}$. In both the NDF and MTL data sets we used, we also found that approximate measure of an outer scale to the edge of the boundary layers is valid under APG, FPG and SFPG conditions in addition to ZPG. Therefore we utilized the outer scale $ \ell \equiv \Delta_{1.25}=1.25 \delta_{99}$, and $\beta \equiv (-\Delta_{1.25}/\tau_{w}) \cdot (\partial P_{e}/\partial x)$.  We also define $Y \equiv y^+ /Re_\tau$ to obtain the following from (\ref{eq:010}):
\begin{equation}\label{eq:012}
{U_x}^+_{\rm OL}(Y\ll 1) \sim \kappa^{-1}\ln Y + \kappa^{-1}\ln Re_\tau +S_0  Y + B_0 + B_1/Re_\tau. 
\end{equation}
Simplifying, we obtain:
\begin{equation}\label{eq:011}
{U_x}^+_{\rm OL}(Y\ll 1) = \kappa^{-1}\ln Y + \kappa^{-1}\ln Re_\tau +S_0  Y + B_0 + {\rm  H.O.T.}
\end{equation}

\begin{figure}
      \centering
      \includegraphics[width=0.85\textwidth]{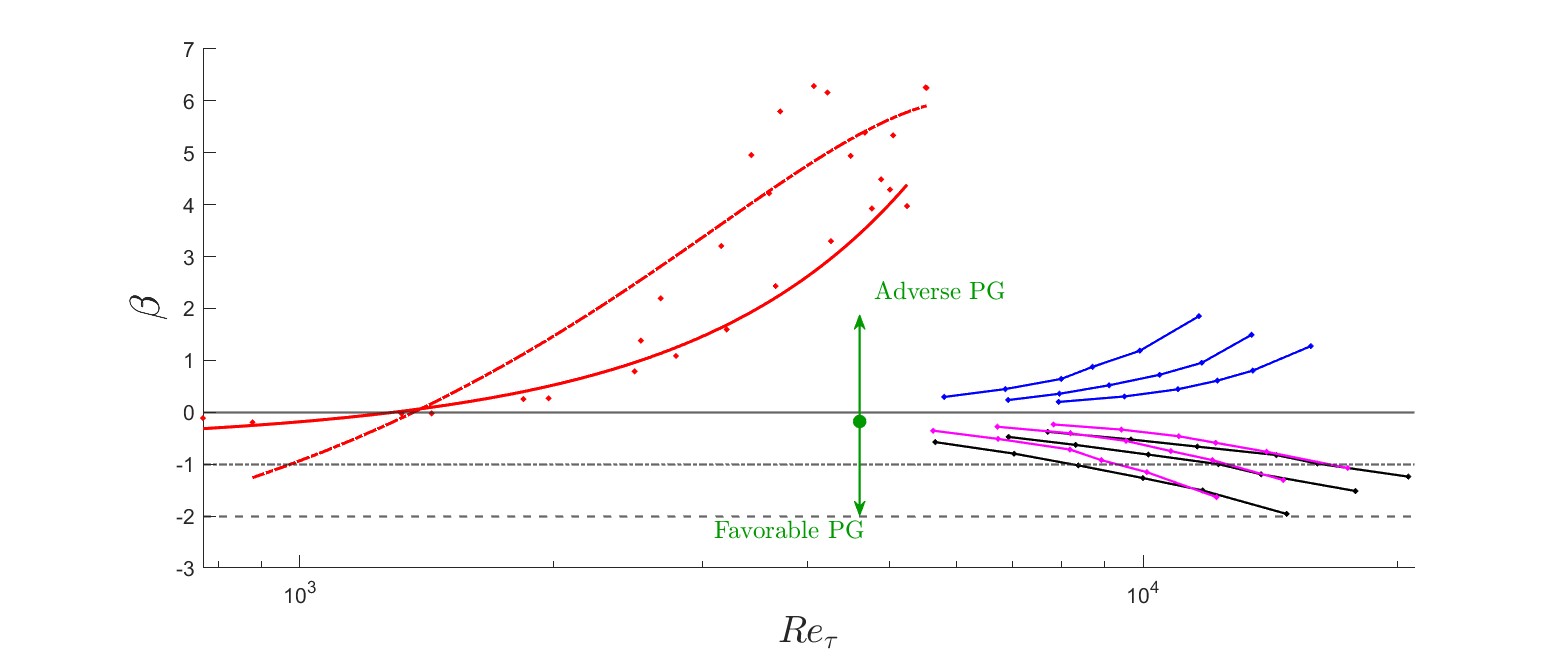}
      \caption{Reynolds-number dependence of pressure-gradient parameter, $\beta$, in boundary layers of data sets used; APG data in red dots with second-order best fit for two configurations in MTL by \cite{sanm20}, ``mildly increasing" APG (dotted red line) and ``approximately constant" APG (red line); data from NDF by \cite{iutam100} for three configurations at six streamwise locations and three freestream velocities, FPG in magenta, SFPG in black and APG in blue; black lines depict ZPG (solid) fully developed channel (dotted) and fully developed pipe (dashed) conditions.}
       \label{fig:fig2}
\end{figure}

With increasing Reynolds number, the higher-order terms (H.O.T.) starting with $B_1/Re_\tau$ may be ignored. The pressure-gradient parameter defined using the outer scale $\Delta_{1.25}$ calculated for all profiles of boundary layers from the NDF and MTL data sets are displayed in figure~\ref{fig:fig2}. Note that in figure 3 of \cite{sanm20} the pressure-gradient parameter used was $\beta^*$ utilizing $\delta^*$ as a length scale and yielding conditions of ``approximately constant'' or ``mildly increasing'' APG.

Finally, the commonly used indicator function based on the mean velocity profile for wall-bounded turbulence, $\Xi$, can be obtained from:
\begin{equation}\label{eq:020}
 \Xi = y^+\frac{{\rm d}{U}_x^+}{{\rm d}y^+} = Y\frac{{\rm d}{U}_x^+}{{\rm d}Y}.   
\end{equation}
From equations (\ref{eq:011}) and (\ref{eq:020}), one can obtain the equation for $\kappa$ and $S_0$:
\begin{equation}\label{eq:021}
\Xi_{\rm OL} = \kappa^{-1} +S_0  y^+ /Re_\tau = \kappa^{-1} +S_0  Y .
\end{equation}

\section{Determining overlap coefficients}
Applying equation (\ref{eq:020}) to mean-velocity-profile data of any wall-bounded flow in inner or outer distances from the wall, $y^+$ or $Y$ respectively, can result in determining $\kappa$ and $S_0$ by a linear fit corresponding to equation (\ref{eq:021}).  One needs to only select the range in  $y^+$ or $Y$ to fit the linear relations.  In figure 3b of \cite{lee15}, for $Re_\tau \approx 5200$, they applied a linear fit over around $300 < y^+ < 700$ ($0.06 < Y < 0.14$) and obtained $\kappa = 0.384$ and $S_0 = 0$. Instead, if they had applied a linear fit over $0.21 < Y < 0.45$ ($1100 < y^+ < 2300$) they would have obtained $\kappa = 0.41$ and $S_0 = 1.11$, which compare very well with values for channel flows obtained by MN23, {\it i.e.}, $\kappa = 0.417$ and $S_0 = 1.1$. Recently, \cite{forprf} obtained values of $\kappa = 0.51$ and $S_0 = 2.6$ also for channel flows computed with two different resolutions using this approach over $0.3 < Y < 0.6$ for $Re_\tau = 550$. The value of $B_0 = 6.9$ was then obtained using the already determined values of $\kappa$ and $S_0$ by minimizing the differences between the velocity profile data to equation (\ref{eq:011}).

\begin{figure}
    \centering
    \begin{minipage}{0.49\textwidth}
        \centering
        \includegraphics[width=\linewidth]{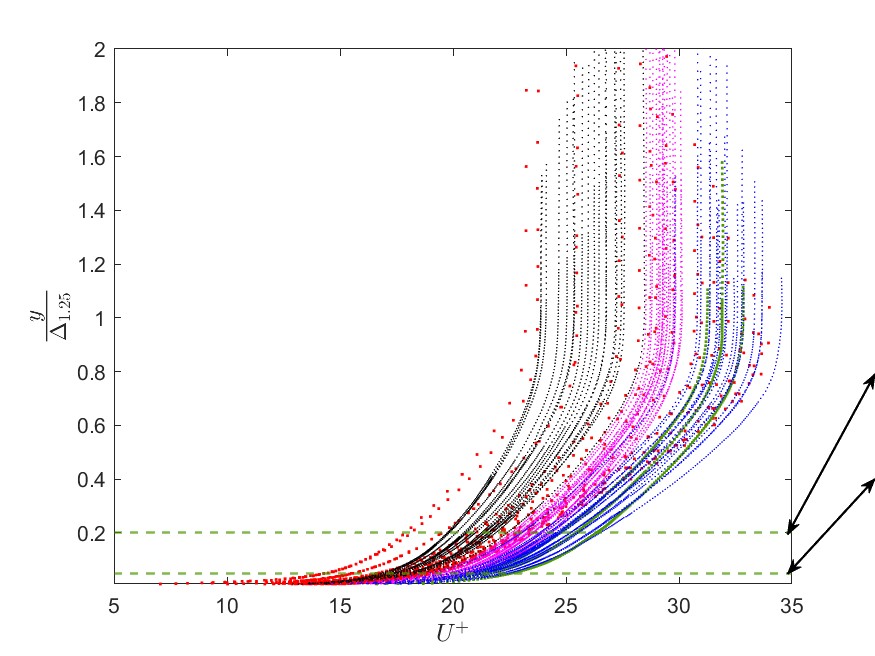}
    \end{minipage}
    \begin{minipage}{0.49\textwidth}
        \centering
        \includegraphics[width=\linewidth]{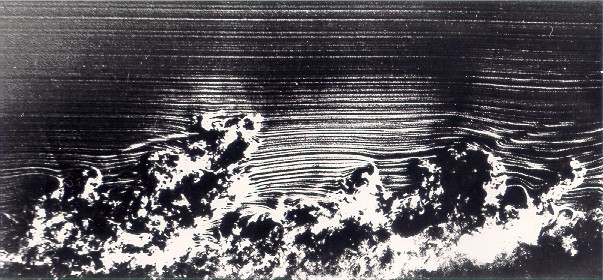}
    \end{minipage}
    \caption{(Left) Inner-scaled mean velocity profiles plotted on linear scales against $y/\Delta_{1.25}$ for boundary layers of data sets used, FPG (magenta), SFPG (black), ZPG (green), and APG (blue) from NDF, and APG (red) from MTL; dashed green lines at $y/\Delta_{1.25} = 0.05~\&~0.2$ correspond to dashed green lines in Figure ~\ref{fig:fig4}; (Right) smoke-wire visualization of ZPG boundary layer in NDF at $Re_\theta$ around $1,800$ with arrows corresponding to approximate locations of $y/\Delta_{1.25} = 0.05~\&~0.2$.}
 \label{fig:fig3}
 \end{figure}

\begin{figure}
      \centering
      \includegraphics[width=0.7\textwidth]{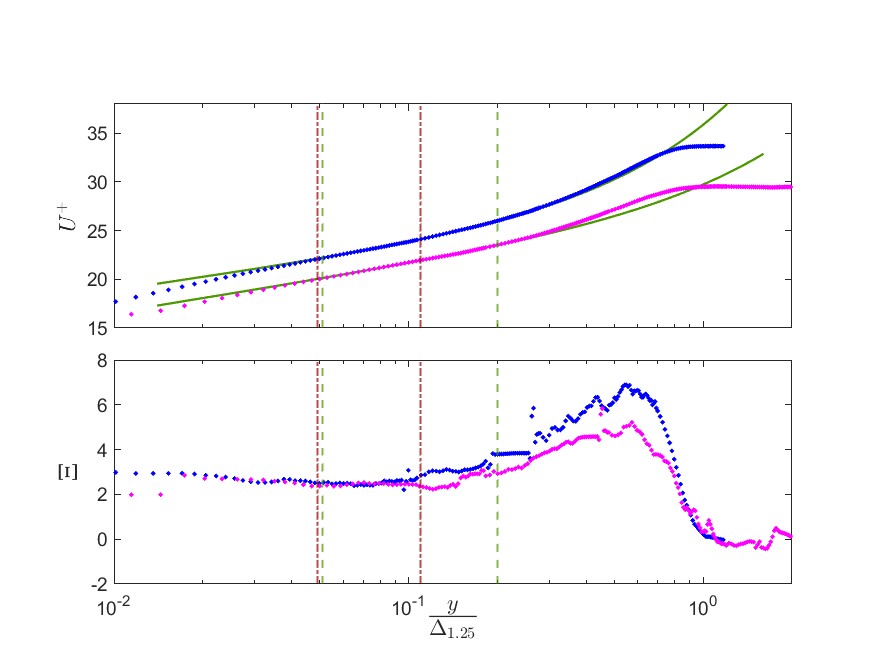}
      \caption{(Top) Inner-scaled mean velocity profile data plotted against logarithmic-scaled $y/\Delta_{1.25}$ for two boundary layers from NDF, for FPG (magenta) and APG (blue); green lines represent the log+lin fit of the overlap region between two vertical dashed green lines at $y/\Delta_{1.25} = 0.05~\&~0.2$, and vertical dashed red lines correspond to $y/\Delta_{1.25} = 0.05~\&~0.11$, which is consistent with MN23; (Bottom) Indicator function, $\Xi$ plotted against same scale of top part.}
       \label{fig:fig4}
\end{figure}

This approach requires highly accurate and closely spaced data for the velocity profiles to apply the derivatives of the indicator function $\Xi$. This is often the case from direct-numerical-simulation (DNS) data, and to some degree in the boundary-layer data from the NDF, where over $200$ data points with equal physical spacing were measured.  Typical wall-bounded-flow measurements, including in boundary layers, collect the measurements using points with logarithmic physical spacing, {\it i.e.}, with increasing physical separation away from the wall. This has been a tradition based on an expected ``log law" in the overlap region.  The MTL data set was obtained in this way with only $45$ to $75$ measurement points in the wall-normal direction. The profiles of figure~\ref{fig:fig3} display the measured points and demonstrate the contrast in spacing between MTL (red) and NDF (blue, green, magenta and black). While determining the indicator function $\Xi$ from the MTL data was attempted using several ways, it was found to be not reliable and produced noisy results.  On the other hand, two examples of $\Xi$ obtained from NDF profiles are shown in figure~\ref{fig:fig4}.  To process both data sets using the same approach and algorithms a different method was required. However, the $\Xi$ profiles from all of the NDF data were very important to select the range of $Y$ to use for the overlap region. Based on the experience from MN223 and \cite{forprf}, it is found that fitting in the outer-scaled distances $Y$ from the wall is more representative of the overlap region than by using inner-scaled distances $y^+$, over wider ranges of $Re_\tau$.  It also helps in avoiding the wake region especially for APG conditions, and other pressure-gradient boundary layers, including ZPG.

The approach used here is different from the one for channel and pipe flows described above as we are retrieving all three parameters of equation (\ref{eq:011}) at the same time, by utilizing all profiles in $U^{+}$ versus Y (=~$y/\Delta_{1.25}$).
After choosing the limits of the overlap region in Y, we compute the best combination of parameters using nonlinear least-squares curve fitting. We fit our function to the data by adjusting the parameters of the function to minimize the sum of squared differences between the observed and predicted values. This optimization involves minimizing the sum of squared residuals. 

Given the set of data points ($Y_{i}$, $U^{+}_{i}$) and the model function $f(\kappa,~S_{0},~B_{0},~Y_{i})$, represented by equation (\ref{eq:011}), we approximate the combination of parameters that minimizes the difference between the equation and the data for each mean velocity profile: $\forall N \in \mathbb{N}^{*}$, $(\kappa_{\rm opt},~S_{0_{\rm opt}},~B_{0_{\rm opt}})=\underset{\kappa,~S_{0},~B_{0}}{\rm argmin}\sum_{n=1}^{N} (U^{+}_{i} - f(\kappa,~S_{0},~B_{0},~Y_{i}))^{2}$. This optimization algorithm, with an accuracy of the order of $10^{-8}$, leads to the final values of the three parameters of the log+lin overlap for each velocity profile, using $Y_{\rm min}=0.05$ and $Y_{\rm max}=0.2$.

\section{Discussion and conclusion}
\begin{figure}
      \centering
      \includegraphics[width=0.98\textwidth]{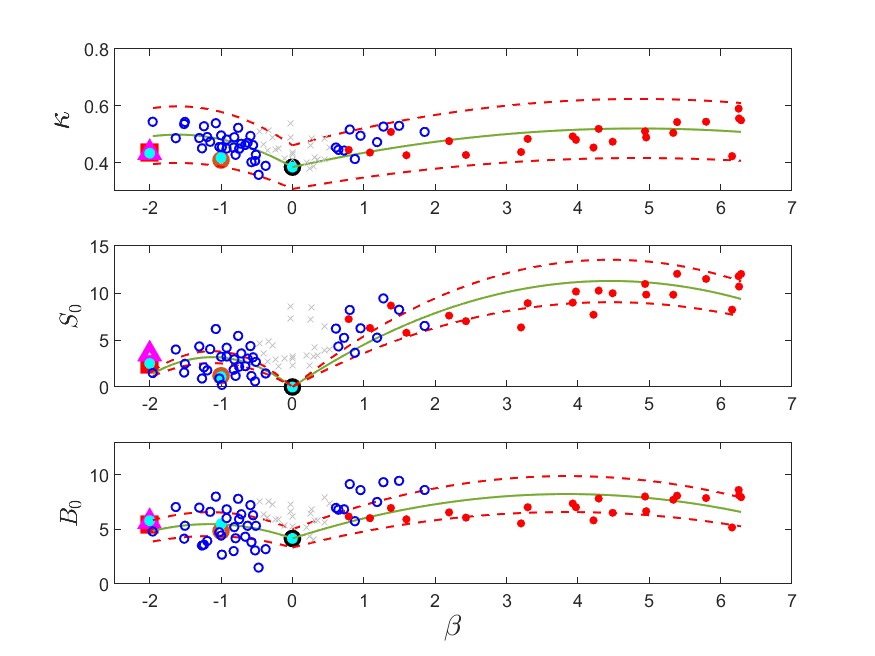}
      \caption{Best-fit overlap parameters, $\kappa$, $S_0$ and $B_0$ based on equation (\ref{eq:011}) plotted against pressure-gradient parameter, $\beta$, for all data sets of Figure 3; NDF in open blue circles and MTL in closed red circles, with ZPG from NDF and \cite{sam18} in large black open circle; parameter values from ~\cite{mon23} in closed cyan circles; parameter values from DNS of \cite{forprf} for channel (large brown open circles) and pipe (large red open square); recent experiments in CICLoPE Pipe in magenta triangle; green lines are second-order best fits of all figure data with dashed red line representing 20\% deviations from fit (see Table 1); faint grey crosses represent profiles with high fit uncertainty. }
       \label{fig:fig5}
\end{figure}

\begin{figure}
      \centering
      \includegraphics[width=0.9\textwidth]{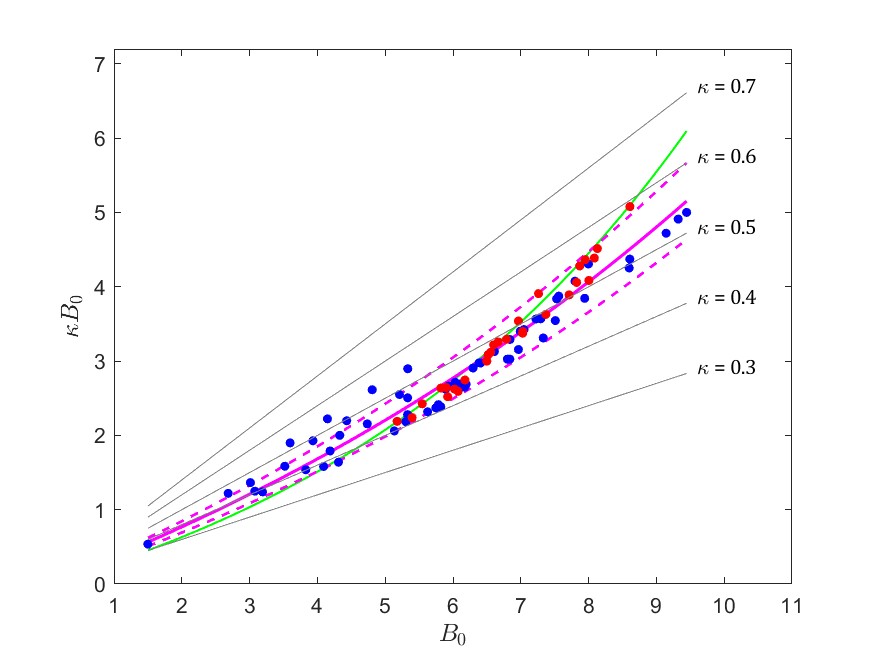}
      \caption{Product of $\kappa B_0$ plotted against $B_0$ to test universality of K\'arm\'an coefficient against lines of constant $\kappa$ based on best-fit overlap parameters from equation (\ref{eq:011}); NDF data in blue dots and MTL data in red dots; magenta line is best fit according to equation (\ref{eq:002}) with $10\%$ deviation indicated by dashed lines; green curve is similar fit of a wide range of data using composite profiles by \cite{variations} for a log-only overlap region, shown in equation (\ref{eq:003}).}
       \label{fig:fig6}
\end{figure}

With the approach described in the last part of the previous section, a parametric evaluation of the range of $Y$ to extract the parameters of the log+lin overlap from the mean velocity profile data was conducted using the NDF data first and then extended to the MTL profiles to arrive at a single range for all data.  The experience from MN23, especially for ZPG, was a good starting point. For the lower limit $Y_{\rm OL,min}$ we arrived at a value of 0.05 to accommodate the lower range of $Re_\tau$'s in the MTL data. For the upper limit of the range $Y_{\rm OL,max}$, we tested several values in the range $0.11 < Y < 0.4$ (compare to figure 7 of MN23). Based on both sets of data, and in order to avoid including parts of the wake, which varies with pressure gradient, we again converged on $Y_{\rm OL,max} = 0.2$.  Two ranges are indicated in figure~\ref{fig:fig4} with dashed lines representing $Y_{\rm OL,min} = 0.05$ and $Y_{\rm OL,max} = 0.11$ in red, and $Y_{\rm OL,min} = 0.05$ and $Y_{\rm OL,max} = 0.2$ in green.  The resulting log+lin overlap fit based on equation (\ref{eq:021}) is shown in solid curves, also in green, with the mean velocity profiles corresponding to an FPG case and an APG case.  Note that for the FPG case with the smaller wake, as indicated by the $\Xi$ curves, the profile data are above the log+lin overlap fit beyond $Y > 0.3$. Further systematic study of the best choice may lead to an ultimate best value of $Y_{\rm OL,max}$ and its possible dependence on pressure gradient, but we are satisfied with the $0.2$ value especially in light of the results presented in figures~\ref{fig:fig5} and \ref{fig:fig6}. 

Boundary layers with small pressure gradients $-0.5 < \beta < 0.5$, including ZPGs, are the most sensitive to the choice of $Y_{\rm OL,max}$ for the accurate extraction of the log+lin overlap parameters and in particular $S_0$, because of the influence of the wake part.  These profiles come from a mixture of NDF cases and the MTL cases, with the fewer data points in the profile, and the resulting parameters are included in figure~\ref{fig:fig5} with grey crosses. The discrimination between the log+lin overlap region and the start of the outer wake region requires closely spaced measurement points.

On the left part of figure~\ref{fig:fig3} we included the selected values  $Y_{\rm OL,min} = 0.05$ and $Y_{\rm OL,max} = 0.2$ on the velocity profiles also in green dashed lines. The arrows starting from those values of $Y = y/\Delta_{1.25}$ to the upstream edge of the flow-visualization image in the right part of the figure are intended to point the approximate corresponding distances from the wall at the bottom of the image.

\begin{table}
  \centering
  \begin{tabular}{c c}
    \hline
    \textbf{Favorable Pressure Gradient} & \textbf{Adverse Pressure Gradient} \\
    \\
    $\kappa_{-} = - 0.044~\beta^{2} - 0.142~\beta + 0.384$ & $\kappa_{+} = - 0.006~\beta^{2} + 0.056~\beta + 0.384$ \\ \\
    $S_{0_{-}} = - 2.516~\beta^{2} - 5.652~\beta$ & $S_{0_{+}} = - 0.569~\beta^{2} + 5.068~\beta$ \\
    \\
    $B_{0_{-}} = - 0.974~\beta^{2} - 2.261~\beta + 4.17$ & $B_{0_{+}} = - 0.275~\beta^{2} + 2.114~\beta + 4.17$ \\
    \hline
    \end{tabular}
  \caption{Equations for second-order best fits of overlap parameters, $\kappa$, $S_0$ and $B_0$,\\for NDF and MTL Data}
  \label{tab:table}
\end{table}

All the data for best-fit values of $\kappa$, $S_0$ and $B_0$ for each boundary-layer velocity profile from both datasets are included in figure~\ref{fig:fig5}. Furthermore, data from MN23, \cite{sam18} and \cite{forprf} are added to the figure. It is important to point out that MN23 defined their pressure-gradient parameter $\beta$ without a minus sign so pipe and channel flows were represented by positive values of $\beta$. All the data in each part of figure~\ref{fig:fig5} were used to fit second-order polynomials, using an algorithm to minimize the differences with an accuracy of the order of $10^{-3}$, resulting in the summary of Table~\ref{tab:table}.

A correlation to test the universality of the K\'arm\'an coefficient was developed by \cite{variations} and is shown by their equation (13), which was based on a log-only overlap region:
\begin{equation}\label{eq:003}
   \kappa B = 1.6~[e^{0.1663 B} - 1].
\end{equation}
For a constant or universal K\'arm\'an constant, $\kappa$, a linear relation between $\kappa B$ and $B$ needs to exist.  In figure~\ref{fig:fig6} we test the universality of $\kappa$ based on the log+lin overlap region and plot all the data from both sets. Lines of various values of constant $\kappa$ are included for reference and none of them represent the entire data sets or segments of them with any reasonable accuracy. Fitting the data extracted using the log+lin overlap by a similar relation to  equation (\ref{eq:011}) we find, using an algorithm to minimize the differences with an accuracy of the order of $10^{-3}$, slightly different coefficients in the relation:
\begin{equation}\label{eq:002}
  \kappa B_0 = 4.05~[e^{0.087 B_0} - 1] .
\end{equation}

Based on the results of figure~\ref{fig:fig5} we conclude that when using the log+lin overlap given by (\ref{eq:011}), no history effects are observed in the overlap region in either of the two data sets used here or between them, over a very wide range of Reynolds numbers and adverse- and favorable-pressure gradients.  We interpret this behaviour as quasi equilibrium of the overlap between outer and inner parts when the new understanding of the overlap region of \cite{mon23} is adopted. In the literature, some history effects have been observed in pressure-gradient boundary layers but the well-documented ones~\citep{harun,bobke} are reflected primarily in Reynolds stresses or the wake part of the mean flow.  We only used the approximately constant and mildly increasing pressure gradient cases from MTL. \cite{sanm20} data included a ``strongly increasing'' case with $\beta^*$  up to 2.38. The validity of our conclusions on the history effects will need to be examined for a wider range of $\beta$ as defined here with an outer length scale. 

\section*{Acknowledgements}
VB and HN acknowledge the ongoing support of the John T. Rettaliata Chair of Mechanical and Aerospace Engineering.  RV acknowledges the financial support from ERC grant no. `2021-CoG-101043998, DEEPCONTROL'. Views and opinions expressed are however those of the author(s) only and do not necessarily reflect those of European Union or European Research Council. Neither the European Union nor granting authority can be held responsible for them.

\section*{Declaration of interests}
The authors report no conflict of interest.

\bibliographystyle{jfm}
\bibliography{jfm-main}

\end{document}